\documentclass[12pt]{article}

\usepackage[reqno]{amsmath}
\usepackage{mathrsfs}
\usepackage{amssymb}

\usepackage{rotating} %for landscape format

\usepackage{bbm}
\usepackage{array}
\usepackage{float}
\usepackage{color}
\usepackage{graphicx}% Include figure files
\usepackage{a4}
\usepackage{a4wide}
\usepackage{wasysym}
\def\gs{\mathrel{
   \rlap{\raise 0.511ex \hbox{$>$}}{\lower 0.511ex \hbox{$\sim$}}}}
\def\ls{\mathrel{
   \rlap{\raise 0.511ex \hbox{$<$}}{\lower 0.511ex \hbox{$\sim$}}}}

\newcommand{\be}{\begin{eqnarray}}
\newcommand{\ee}{\end{eqnarray}}
\newcommand{\beq}{\begin{equation}}
\newcommand{\eeq}{\end{equation}}

\def\gsim{\:\raisebox{-0.5ex}{$\stackrel{\textstyle>}{\sim}$}\:}

\newcommand{\eps}{\mbox{$\epsilon$}}
\newcommand{\Om}{\mbox{$\Omega$}}

\begin{document}

%For feynmf-Package:
\setlength{\unitlength}{1mm}

\begin{titlepage}
\title{\vspace*{-2.0cm}
%\hfill {\small hep--ph/xxxxxx}\\[20mm]
\bf\Large
Radiative Inflation and Dark Energy
\\[5mm]\ }

\author{
Pasquale Di Bari$^a$\thanks{email: \tt P.Di-Bari@soton.ac.uk}~~,~~
Stephen F.\ King$^a$\thanks{email: \tt S.F.King@soton.ac.uk}~~,~~
Christoph Luhn$^a$\thanks{email: \tt Christoph.Luhn@durham.ac.uk}~~,~~\\
Alexander Merle$^{bc}$\thanks{email: \tt amerle@kth.se}~~,~~and~~
Angnis Schmidt-May$^b$\thanks{email: \tt angnis.schmidt-may@fysik.su.se}
\\ \\
{\normalsize $^a$ \it School of Physics and Astronomy, University of Southampton,}\\
{\normalsize \it Southampton, SO17 1BJ, United Kingdom}\\
\\
{\normalsize $^b$ \it Max-Planck-Institut f\"ur Kernphysik,}\\
{\normalsize \it Postfach 10 39 80, 69029 Heidelberg, Germany}\\
\\
{\normalsize $^c$ \it Department of Theoretical Physics, School of Engineering Sciences,}\\
{\normalsize \it Royal Institute of Technology (KTH) -- AlbaNova University Center,}\\
{\normalsize \it Roslagstullsbacken 21, 106 91 Stockholm, Sweden}
}
\date{\today}
\maketitle
\thispagestyle{empty}

\begin{abstract}
\noindent
We propose a model based on radiative symmetry breaking that combines inflation with Dark Energy and is consistent with the WMAP 7-year regions. The radiative inflationary potential leads to the prediction of a spectral index $0.955 \lesssim n_S \lesssim 0.967$ and a tensor to scalar ratio $0.142 \lesssim r \lesssim 0.186$, both consistent with current data but testable by the Planck experiment. The radiative symmetry breaking close to the Planck scale gives rise to a pseudo Nambu-Goldstone boson with a gravitationally suppressed mass which can naturally play the role of a quintessence field responsible for Dark Energy. Finally, we present a possible extra dimensional scenario in which our model could be realised.
\end{abstract}

\end{titlepage}

%%%%%%%%%%%%%%%%%%%%%%%%%%%%%%%%%%%%%%%%%%%%%%%%%%%%%%%%%%%%%%%%%%%%%%
\section{\label{sec:Introduction}Introduction}
%%%%%%%%%%%%%%%%%%%%%%%%%%%%%%%%%%%%%%%%%%%%%%%%%%%%%%%%%%%%%%%%%%%%%%

Although modern cosmology seems to require both inflation and Dark Energy there are relatively few models which attempt to unify these two ideas ~\cite{Kinney:2009vz,Copeland:2006wr}. One of the most interesting attempts to achieve such a unification has relatively recently been discussed~\cite{Rosenfeld:2005mt}, based on the earlier `schizon model'~\cite{Hill:1988bu,Frieman:1991tu,Frieman:1995pm}. This model was, however, essentially based on   $\varphi^4$ chaotic inflation, which was significantly threatened by the  Wilkinson Microwave Anisotropy Probe (WMAP) 5-year data~\cite{Hinshaw:2008kr} (if not ruled out).\footnote{Note, however, that the situation of the model from Ref.~\cite{Rosenfeld:2005mt} looks much better in the case of small field inflation~\cite{Rosenfeld:2006hs}.} However, the model has some nice features as, e.g., naturally generating a pseudo Nambu-Goldstone boson (PNGB), which receives a potential via gravitational effects~\cite{Kallosh:1995hi} and can then be used as quintessence field. Other attempts can provide a better match to the data by invoking hybrid inflation, with~\cite{Masso:2006yk} or without~\cite{Gong:2005bn} using a PNGB as quintessence field (see Refs.~\cite{Kaloper:2005aj,Dutta:2006cf} for an extensive discussion of that subject).

In this letter we propose a simple new model which can overcome the difficulties of $\varphi^4$ chaotic inflation but which can also lead to a PNGB quintessence field. The new model is based on the idea of a massive complex scalar field whose mass squared is driven negative close to the Planck scale by radiative effects, leading to a model of Radiative Inflation and Dark Energy (RIDE). The complex scalar field $\Phi$ has a potential which is invariant under a global $U(1)$-symmetry which, in turn, is broken by radiative effects leading to an almost massless PNGB. Close to the Planck scale the inflaton field $\tilde \eta=\sqrt{2} |\Phi|$ rolls slowly down a simple potential that resembles $\varphi^2$ chaotic inflation for high field values. After inflation, however, it settles at its minimum, thereby breaking the global $U(1)$ and generating the Nambu-Goldstone boson which would be massless in the absence of gravitational effects.
Including gravitational effects generates a potential for the PNGB, so that it can then play the role of the quintessence field. The RIDE model leads to interesting predictions for inflation which are fully consistent with WMAP 7-year data~\cite{Komatsu:2010fb} but which allow the model to be ruled out or confirmed by the Planck experiment.

Radiative corrections have been studied before in both, the context of inflation (see, e.g., Refs.~\cite{Dvali:1994ms,Kinney:1995xv,Stewart:1996ey,Stewart:1997wg,Covi:1998mb,Covi:1998jp,Covi:1998yr,German:1999gi,Covi:2004tp,Bugaev:2008bi}) and in the context of quintessence (see, e.g., Refs.~\cite{Kolda:1998wq,Uzan:1999ch,Bartolo:1999sq,Brax:1999yv,Doran:2002bc,Onemli:2002hr,Garny:2006wc}). In contrast to previous studies, however, the radiative corrections here 
are responsible for symmetry breaking via a scalar mass squared being driven negative at a high scale
close to the Planck scale, allowing us to relate inflation to quintessence.

The remainder of this letter is organised as follows. After introducing the model in Sec.~\ref{sec:Model}, we perform analyses of inflation and quintessence in Secs.~\ref{sec:Inflation} and~\ref{sec:Quintessence}, respectively. Finally, we sketch a possible scenario in which our model could be realised in Sec.~\ref{sec:SUGRAModel}, before concluding in Sec.~\ref{sec:Conclusions}. More details on the effects of radiative corrections on the potential can be found in the Appendix.

%%%%%%%%%%%%%%%%%%%%%%%%%%%%%%%%%%%%%%%%%%%%%%%%%%%%%%%%%%%%%%%%%%%%%%
\section{\label{sec:Model}The Model}
%%%%%%%%%%%%%%%%%%%%%%%%%%%%%%%%%%%%%%%%%%%%%%%%%%%%%%%%%%%%%%%%%%%%%%

The model is based on a complex scalar field $\Phi=\frac{1}{\sqrt{2}} \tilde \eta e^{i\phi/f}$ (with $f=\langle \tilde \eta \rangle$), whose potential in the absence of radiative corrections has
a simple quadratic form, $V_0\approx M^2 \Phi^\dagger \Phi$.\footnote{Note that, when trying to relate such a potential to a concrete particle physics model, it has to be verified that a possible $(\Phi^\dagger \Phi)^2$-term is absent or at least suppressed. Indeed, such a framework can be realised in certain scenarios, see Sec.~\ref{sec:SUGRAModel} for an example.} The basic idea is that radiative corrections then drive the mass squared negative at some scale $\Lambda$ not too far below the Planck scale. This radiative symmetry breaking mechanism is perhaps most familiar in the minimal supersymmetric standard model (MSSM) where top and stop loops drive the Higgs mass squared negative at the TeV scale~\cite{Ibanez:1982fr}, but has been recently used elsewhere in different contexts where a mass squared is driven negative at a much higher scale~\cite{Varzielas:2006ma,Howl:2009ds}.
Such a radiative potential may be parametrised as in~\cite{Varzielas:2006ma,Howl:2009ds},
\begin{equation}
 V\approx M^2 \Phi^\dagger \Phi \ln \left( \frac{\Phi^\dagger \Phi}{\Lambda^2} \right)= \frac{M^2}{2} \tilde \eta^2  \ln \left( \frac{\tilde \eta^2}{2\Lambda^2} \right).
 \label{eq:inf_pot}
\end{equation}
This leads to a vacuum expectation value (VEV) of $f=\sqrt{\frac{2}{e}}\Lambda$ for $\tilde \eta$. In such a potential, inflation can completely take place in a region where $\tilde \eta \gg \Lambda$, in which the $\ln$-term in Eq.~\eqref{eq:inf_pot} is well behaved and the inflaton field $\tilde \eta$ only feels a potential that is very similar to the one used for quadratic inflation.\footnote{Note that we concentrate on the corrections due to the renormalization group evolution, just as done in Refs.~\cite{Stewart:1996ey,Stewart:1997wg,Covi:2004tp}, which is the dominant contribution of the Coleman-Weinberg correction~\cite{Coleman:1973jx} in the case of broken supersymmetry~\cite{Lyth:2007qh}. See also Ref.~\cite{Shafi:2006cs} for experimental constraints on such corrections.}

\begin{figure}[t]
\centering
\begin{tabular}[h]{lr}
\includegraphics[width=8cm]{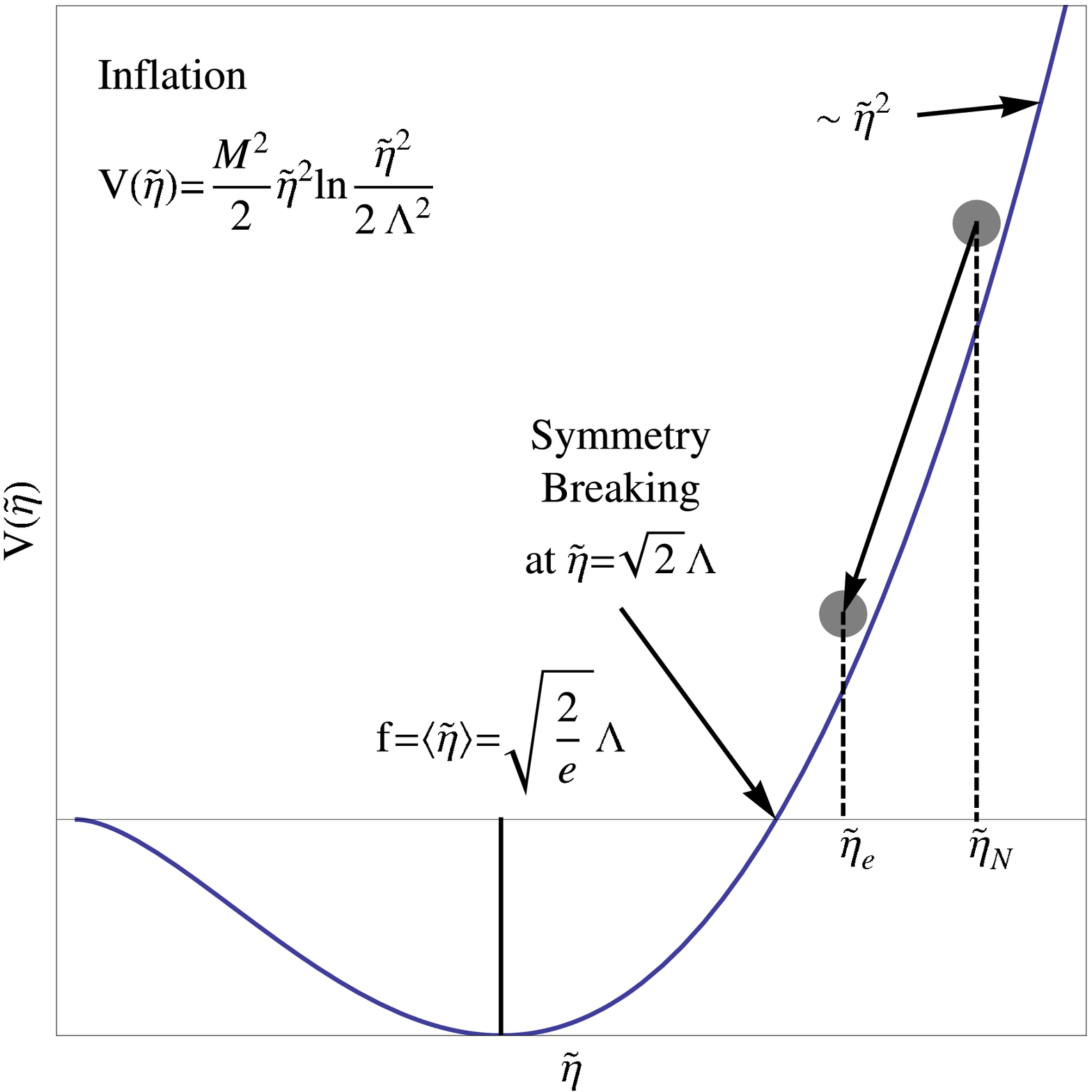}  &
\includegraphics[width=8cm]{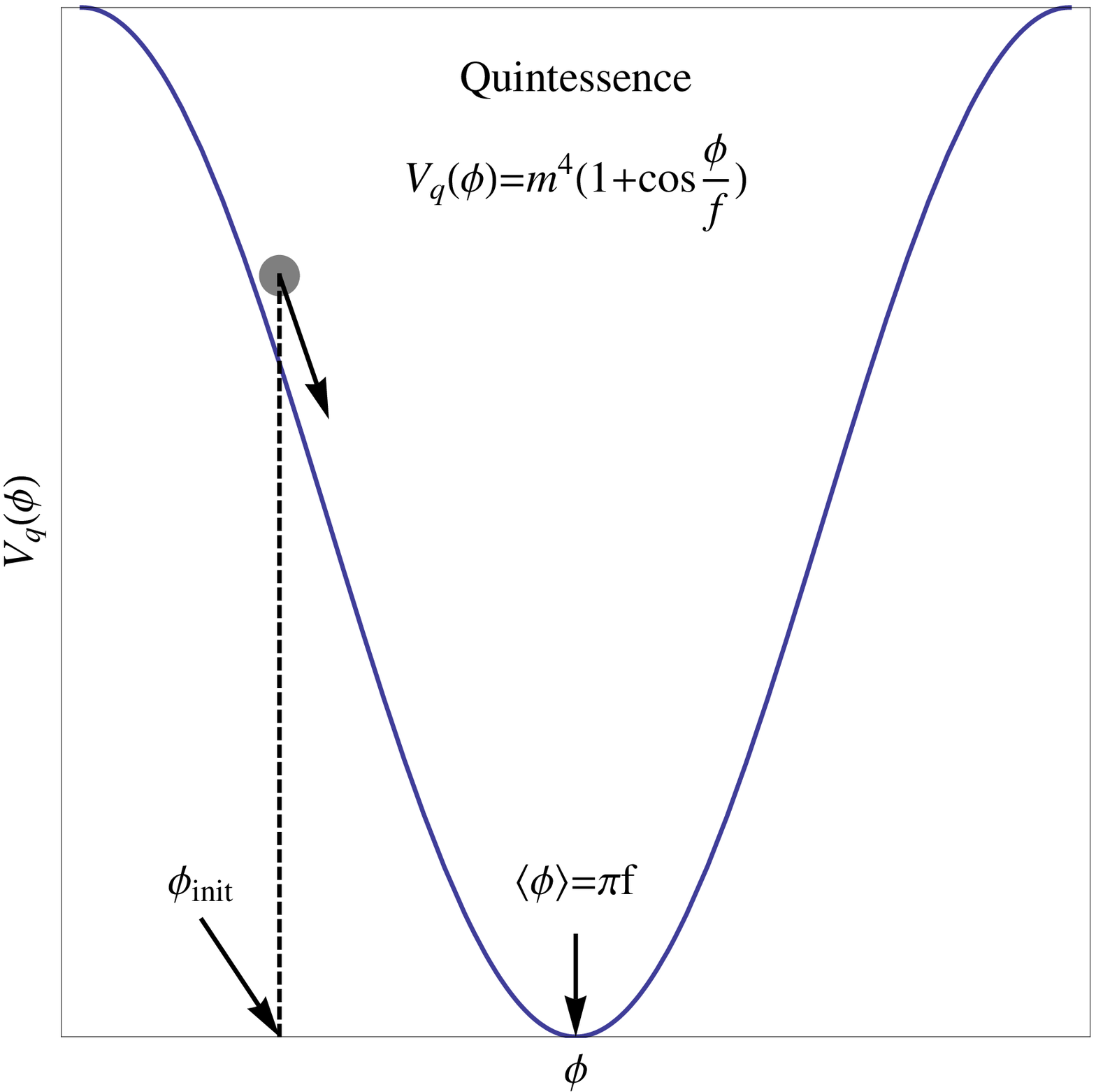}
\end{tabular}
\caption{\label{fig:potentials}The shape of the inflationary potential describing the inflaton
$\tilde \eta$ (left panel) and the potential describing the quintessence field $\phi$ (right panel).
Both fields originate from the complex scalar field $\Phi=\frac{1}{\sqrt{2}} \tilde \eta e^{i\phi/f}$ described
by the potential in Eq.~\eqref{eq:inf_pot}.}
\end{figure}

Later on, the field will settle at its VEV. As the potential is symmetric under a global $U(1)$, either imposed or accidental, the VEV will break this global symmetry, thereby generating a massless Nambu-Goldstone boson $\phi=f \arg (\Phi)$. This field has no mass term and in fact no potential at all. The original $U(1)$ symmetry of $\Phi$ translates to a shift symmetry $\frac{\phi}{f} \to \frac{\phi}{f} + \alpha$, with $\alpha$ being a continuous real parameter. However, the continuous shift symmetry can be broken by gravitational effects~\cite{Rosenfeld:2005mt,Kallosh:1995hi}, dubbed {\it gravitational instantons}, which, similar to the case of the axion~\cite{Preskill:1982cy,Weinberg:1977ma}, can generate a mass term. Although they break the shift symmetry, these gravitational effects leave invariant a discrete subgroup of transformations, namely those for which $\alpha=2\pi n$, with $n\in \mathbb N$ denoting the winding number of equivalent vacua which one can freely choose also in the presence of gravitational corrections. Hence, any potential $V_q(\phi)$ that is generated by such effects must still be invariant under $\frac{\phi}{f} \to \frac{\phi}{f} + 2\pi n$. In order to have a mass term for $\phi$ in its Taylor expansion, the potential must be an even $2\pi n$-periodic function. The most general such function is a sum of cosines whose arguments are integer multiples of $\frac{\phi}{f}$. It is possible to argue, see~\cite{Sorbo:2008zz}, that the dominant contribution is obtained from the lowest harmonic $\propto \cos \left( \frac{\phi}{f}\right)$, so that the resulting potential
for the quintessence field $\phi$
reads\footnote{Note that, in certain settings, it might be necessary to protect this potential against too large radiative corrections, see Refs.~\cite{Hill:1988bu,Frieman:1991tu}.}
\begin{equation}
 V_q(\phi)=m^4 \left[ 1+\cos \left( \frac{\phi}{f} \right) \right].
 \label{eq:quint_pot}
\end{equation}
Both potentials, $V(\tilde \eta)$ and $V_q(\phi)$, as well as the field dynamics are schematically depicted in Fig.~\ref{fig:potentials}. Note that the dynamics of both sectors can be easily disentangled, as the kinetic term simplifies to
\begin{equation}
 (\partial_\mu \Phi)^* (\partial^\mu \Phi)=\frac{1}{2} (\partial_\mu \tilde \eta)(\partial^\mu \tilde \eta) + \frac{\tilde \eta^2}{2 f^2}(\partial_\mu \phi) (\partial^\mu \phi),
 \label{eq:kin_term}
\end{equation}
with the $\phi$-part being negligible during inflation and $\tilde \eta$ already sitting at its (constant) VEV $f$ during quintessence. Due to this separation of the dynamics of the two fields, we should be safe from potentially dangerous corrections due to (iso-) curvature fluctuations that can appear in multi-field inflation models~\cite{Lalak:2007vi}, since we are practically dealing with a single-field potential.

%%%%%%%%%%%%%%%%%%%%%%%%%%%%%%%%%%%%%%%%%%%%%%%%%%%%%%%%%%%%%%%%%%%%%%
\section{\label{sec:Inflation}Inflation}
%%%%%%%%%%%%%%%%%%%%%%%%%%%%%%%%%%%%%%%%%%%%%%%%%%%%%%%%%%%%%%%%%%%%%%

The inflaton potential of Eq.~\eqref{eq:inf_pot} depends on two parameters, $M$ and $\Lambda$. In this section we show that they can be chosen such as to be consistent with the WMAP 7-year data. Assuming $\Lambda$ close to the Planck scale, we define
\be
\Lambda~=~ k\,M_P ,
\ee
and fix $k$ to take a particular value, like e.g. $0.01$. With this specific choice, all dimensionful quantities, including the parameter $M$, can be expressed in terms of $M_P$ only.

To determine $M$, and subsequently the scalar spectral index $n_S$ as well as the tensor to scalar ratio $r$, it is convenient to start with the slow-roll parameters,
\begin{equation}
 \eps =\frac{M_P^2}{16\pi} \left( \frac{V'}{V} \right)^2=\frac{M_P^2}{4\pi \tilde \eta^2} \left(1 + \frac{1}{L} \right)^2\ {\rm and}\ \eta=\frac{M_P^2}{8\pi} \left[ \frac{V''}{V} -\frac{1}{2} \left( \frac{V'}{V} \right)^2\right]=\frac{M_P^2}{4\pi \tilde \eta^2} \left(\frac{1}{L} - \frac{1}{L^2} \right),
 \label{eq:slow_roll}
\end{equation}
where $L=\ln \left( \frac{\tilde \eta^2}{2\Lambda^2} \right)$. The field value $\tilde \eta_e$ at the end of inflation is calculated numerically by setting $\eps=1$. Note that, in the interesting part of the parameter space, $\tilde \eta_e$ is always very well above $f$, though not necessarily by orders of magnitude. The next quantity we determine is the field value $\tilde \eta_N$, $N$ $e$-folds before the end of inflation. Since there is again no simple approximation, we determine $\tilde \eta_N$ by numerically solving
\begin{equation}
 N~\simeq~\frac{8\pi^2}{M_P^2} \int_{\tilde\eta_e}^{\tilde\eta_N}
 \frac{V(\tilde\eta)}{V^\prime(\tilde\eta)} d\tilde\eta
~=~2\pi \left[ \frac{\tilde \eta_N^2 - \tilde \eta_e^2}{M_P^2} - \frac{2}{e} \left( \frac{\Lambda}{M_P} \right)^2 \left[ {\rm Ei}(1+L_N) - {\rm Ei}(1+L_e)\right] \right],
 \label{eq:eta_N}
\end{equation}
where ${\rm Ei}(z)$ is the exponential integral ${\rm Ei}(z)=-\int_{-z}^{\infty} \frac{e^{-t}}{t} dt$, $L_i=\ln \left(\frac{\tilde \eta_i^2}{2\Lambda^2} \right)$, and $N$ lies within the interval $N\in [46,60]$. Using the so obtained value of $\tilde \eta_N$, the parameter $M$ in Eq.~\eqref{eq:inf_pot} is constrained by the size of the scalar perturbations in the Cosmic Microwave Background (CMB)~\cite{Komatsu:2010fb},
\begin{equation}
 P_\mathcal{R}^{1/2}=\frac{H(\tilde \eta_N)}{M_P \sqrt{\pi \eps(\tilde \eta_N)}}\simeq 4.95\cdot 10^{-5}\, , {\rm with}\ H^2\simeq \frac{8\pi V}{3 M_P^2},
 \label{eq:scalar_pert}
\end{equation}
leading to $M\simeq [10^{-8} M_P, 10^{-7} M_P]$. We have checked that this result is nearly independent of $\Lambda$, which only enters logarithmically. Hence our predictions for inflation are very stable with respect to adjustments of $\Lambda$ which, as we will show later, are necessary to satisfy the constraints coming from the quintessence side.

\begin{figure}[t]
\centering
\includegraphics[width=14cm]{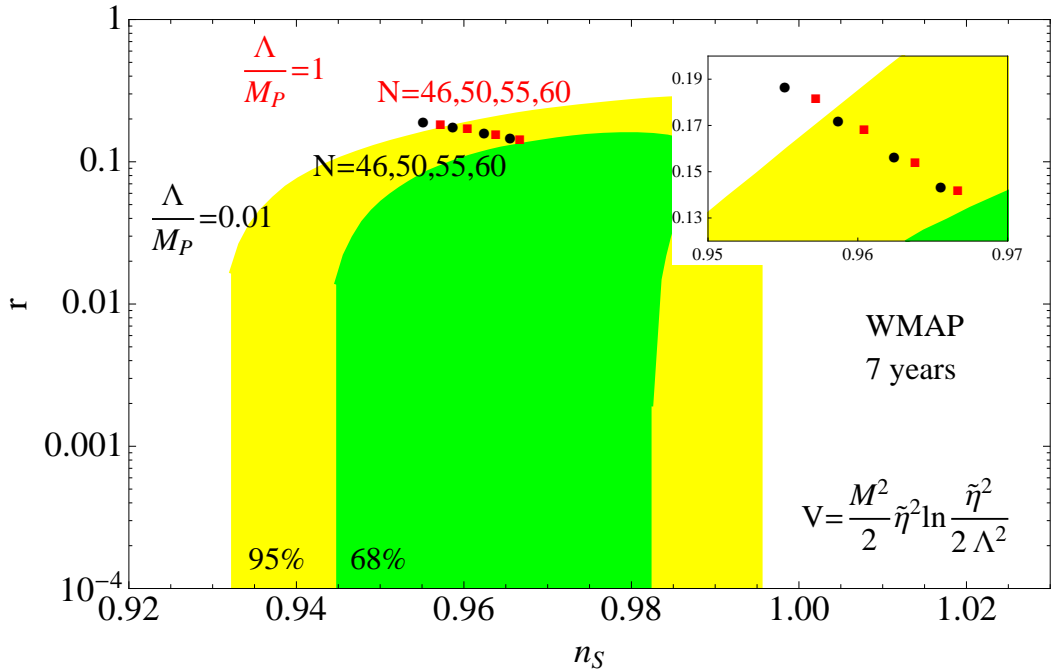}
\caption{\label{fig:inflation}The predictions of the RIDE model for the spectral index $n_S$
and tensor to scalar ratio $r$ as compared to the WMAP 7-year data~\cite{Komatsu:2010fb}, with the inset showing a blow-up of the interesting region. The red squares (black circles) are for $k=1$ ($k=0.01$) for values of $N=46-60$.}
\end{figure}

The above discussion shows how to determine the parameter $M$ for certain values of $k=\frac{\Lambda}{M_P}$ and $N$. With this the potential of Eq.~\eqref{eq:inf_pot} is completely fixed and we can calculate the scalar spectral index, $n_S=1-4 \eps (\tilde \eta_N) + 2 \eta (\tilde \eta_N)$, as well as the tensor to scalar ratio, $r=16\eps (\tilde \eta_N)$, for different values of $k=\frac{\Lambda}{M_P}$ and $N$. The corresponding predictions are in the ranges
\be
0.955 \lesssim n_S \lesssim 0.967~ {\rm and} ~ 0.142 \lesssim r \lesssim 0.186 ,
\ee
and are perfectly consistent with the $95\%$ region of the WMAP 7-year data for $N=50-60$,
as shown in Fig.~\ref{fig:inflation}. Trans-Planckian values of $\Lambda$, which often appear in large field inflation models~\cite{Kinney:2009vz}, would improve the consistency with data even further. Note that, however, such high values are under some dispute~\cite{ArkaniHamed:2003wu}. In any case, we do not need such extreme VEVs, and values of $\Lambda$ around $0.1 M_P$ (or slightly larger) are perfectly fine for our model, as we will see in the next section.

%%%%%%%%%%%%%%%%%%%%%%%%%%%%%%%%%%%%%%%%%%%%%%%%%%%%%%%%%%%%%%%%%%%%%%
\section{\label{sec:Quintessence}Quintessence}
%%%%%%%%%%%%%%%%%%%%%%%%%%%%%%%%%%%%%%%%%%%%%%%%%%%%%%%%%%%%%%%%%%%%%%

The quintessence part of the potential, Eq.~\eqref{eq:quint_pot}, arises from (non-perturbative) gravitational effects, as indicated in Sec.~\ref{sec:Model}. Such corrections induced by gravity, though hard to avoid, are expected to be exponentially suppressed~\cite{Kallosh:1995hi}. Although this might make them sound negligible, in the absence of other corrections, such gravitational corrections will determine the potential for the pseudo Nambu-Goldstone boson,
leading to a suitable quintessence interpretation. The Dark Energy scale $m$ will be determined by
$m^4=e^{-S} M_P^3 f$, where $f=\langle \tilde \eta \rangle$ and $S\sim \pi \frac{M_P^2}{M_{\rm string}^2}$ is a potentially large instanton action~\cite{Kallosh:1995hi,Masso:2006yk,Kaloper:2008fb,SorboTalk}, with $M_{\rm string}$ being the scale of string theory. Assuming $m\sim 10^{-3}~{\rm{eV}}$ the ratio $\frac{M_{P}}{M_{\rm string}}$ is required to be around 10, which is not unreasonable. The message is that, although not giving a prediction, such considerations give at least a motivation for why $m^4$ should be small in the first place.

Assuming the smallness of $m^4 \sim V_q(\phi_0) \sim \rho_{\phi,0}$ to be given, this quantity must be of the order of the current critical density of the Universe, $\rho_{c,0}=\frac{3 H_0^2 M_P^2}{8\pi}$, so that the current Dark Energy fraction $\Om_{\Lambda,0}=\frac{\rho_{\Lambda,0}}{\rho_{c,0}}$ equals $0.728^{+0.015}_{-0.016}$~\cite{Komatsu:2010fb}. A further constraint arises from the requirement that the quintessence field must not have settled at its VEV today, which translates into a bound on its mass, $M_\phi=\frac{m^2}{f}\lesssim 3 H_0$~\cite{Rosenfeld:2005mt,Frieman:1995pm}. Both these conditions lead to a bound on $f$ (and thus also on $\Lambda=\sqrt{\frac{e}{2}}f$) which should be $f\gtrsim 0.1 M_P$~\cite{Rosenfeld:2005mt} or, if one wants to avoid too much tuning, even $f\gtrsim 0.5 M_P$~\cite{Dutta:2006cf}. Similarly as for inflation, one might question values of $f$ too close to the Planck scale, a problem that can be cured by, e.g., invoking extra spatial dimensions~\cite{Pilo:2003gu}. We have analysed the potential in Eq.~\eqref{eq:quint_pot} with $f=M_P/\sqrt{8\pi}$, using an extended version of the {\tt SuperCosmology} package~\cite{Kallosh:2004rs}, where we have also included the cosmological evolution of radiation. This means that we numerically solve the acceleration equation (which is, for a flat Universe, equivalent to the Friedmann equation),
\begin{equation}
 \frac{\ddot{a}}{a}=\dot{H}+H^2=-\frac{4\pi}{3 M_P^2} (\rho_{\rm tot}+3p_{\rm tot}),
 \label{eq:Raychaudhuri}
\end{equation}
where $\rho_{\rm tot}=\rho_{\rm rad}+\rho_{\rm mat}+\rho_{\phi}$ and $p_{\rm tot}=p_{\rm rad}+p_{\rm mat}+p_{\phi}$ are the total energy density and the total pressure. Conveniently, we can immediately insert the known evolutions of radiation and matter,
\begin{equation}
 \rho_{\rm rad}=\frac{\rho_{\rm rad,init}}{a^4},\ p_{\rm rad}=\frac{1}{3} \rho_{\rm rad},\ \rho_{\rm mat}=\frac{\rho_{\rm mat,init}}{a^3},\ p_{\rm mat}=0,
 \label{eq:rad_mat}
\end{equation}
where $a$ is the scale factor. The energy density and the pressure of the quintessence field, however, are only known as functions of $\phi$, $\rho_\phi=\frac{1}{2}\dot{\phi}^2+V_q$ and $p_\phi=\frac{1}{2}\dot{\phi}^2-V_q$. Note that it is often convenient to use the so-called equation of state (EoS) parameter $w$, which is always defined as the ratio between pressure and energy density. For example, $w_{\rm rad}=\frac{1}{3}$ and $w_{\rm mat}=0$. To find the evolution of the quintessence field $\phi$, we have to solve the corresponding equation of motion,
\begin{equation}
\ddot \phi + 3H\dot \phi +V_q'(\phi) = 0 ,
\end{equation}
supplemented by the definitions of the field momentum $P$ and the Hubble parameter $H$,
\begin{equation}
 P = a^3\dot \phi  \ ~ {\rm and}\  H = \frac{\dot{a}}{a}\ .
 \label{eq:DE_evol}
\end{equation}
This gives a total of 4 ordinary first order differential equations to determine the 4 functions $\phi$, $P$, $a$, and $H$. The key ingredient to this system of equations is the quintessence potential, Eq.~\eqref{eq:quint_pot}, which determines the qualitative evolution of the Universe. Note that, however, due to 
$\Gamma=\frac{V_q\,V_q''}{(V_q')^2}\ll 0$ during the slow roll, our potential cannot exhibit a tracking behaviour (which would require $\Gamma >1$)~\cite{Steinhardt:1999nw}, which means that specific initial conditions have to be imposed at the beginning of Big Bang cosmology, i.e., after reheating. Starting with initial values of $\Om_{\rm rad,init}=0.99$, $\Om_{\rm mat,init}=0.01$, and $\Om_{\phi,\rm init}=10^{-11}$ (where the smallness of the latter is related to the tiny value of $m$), we have solved the evolution equations numerically, where we have determined the current time by matching the Dark Energy density parameter to its current value $\Om_{\phi,0}=\Om_{\Lambda,0}$. In order to also hit the other ranges from the WMAP 7-year data at $1\sigma$ or $68\%$~C.L.,  i.e.\ $\Om_{\rm mat,0} = 0.2726 \pm 0.0141$ and $w_{\Lambda,0} =-0.980 \pm 0.053$, as well as a proper age of the Universe $\sim 1/H_0$, with $H_0=70.4^{+1.3}_{-1.4} \frac{\rm km/s}{\rm Mpc}$~\cite{Komatsu:2010fb}, we had to choose $m^4\sim\rho_{c,0}/3$ and an initial field value in the range $0<\phi_{\mathrm{init}}<0.16\cdot 2\pi f$. The initial speed of the field does not influence the cosmological evolution remarkably \cite{Dutta:2006cf}, and we therefore set $\dot{\phi}_{\mathrm{init}}=0$. The constraint for the total EoS parameter of the Universe, $w_{\rm tot,0}=w_{\Lambda,0}\cdot\Om_{\Lambda,0}=-0.713\pm0.041$, was obtained from the WMAP results for $\Om_{\Lambda,0}$ and $w_{\Lambda,0}$, under the assumptions of a flat Universe and negligible radiation.
\begin{figure}[t]
\centering
\begin{tabular}[h]{lr}
\includegraphics[width=7.6cm]{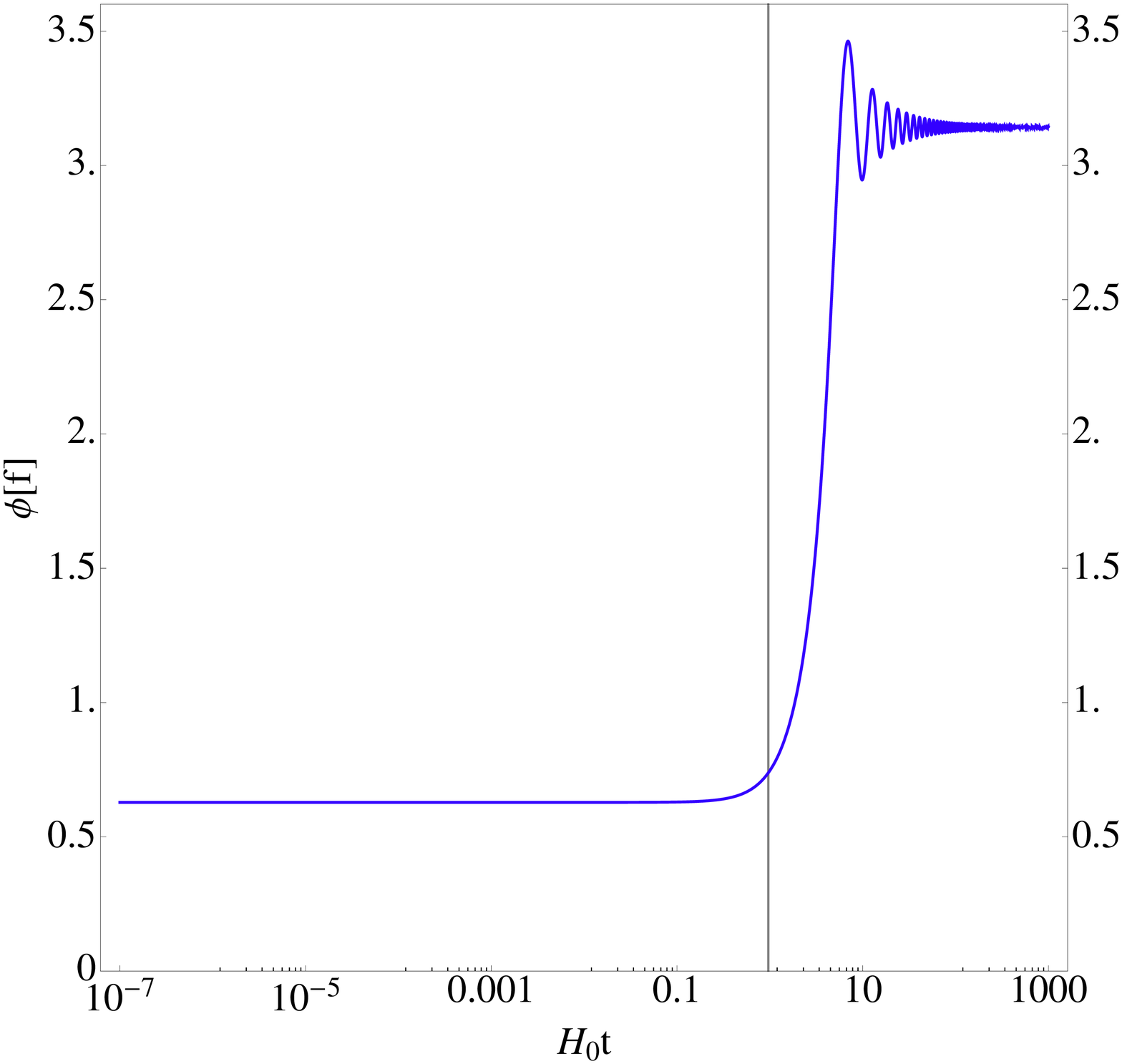} &
\includegraphics[width=8cm]{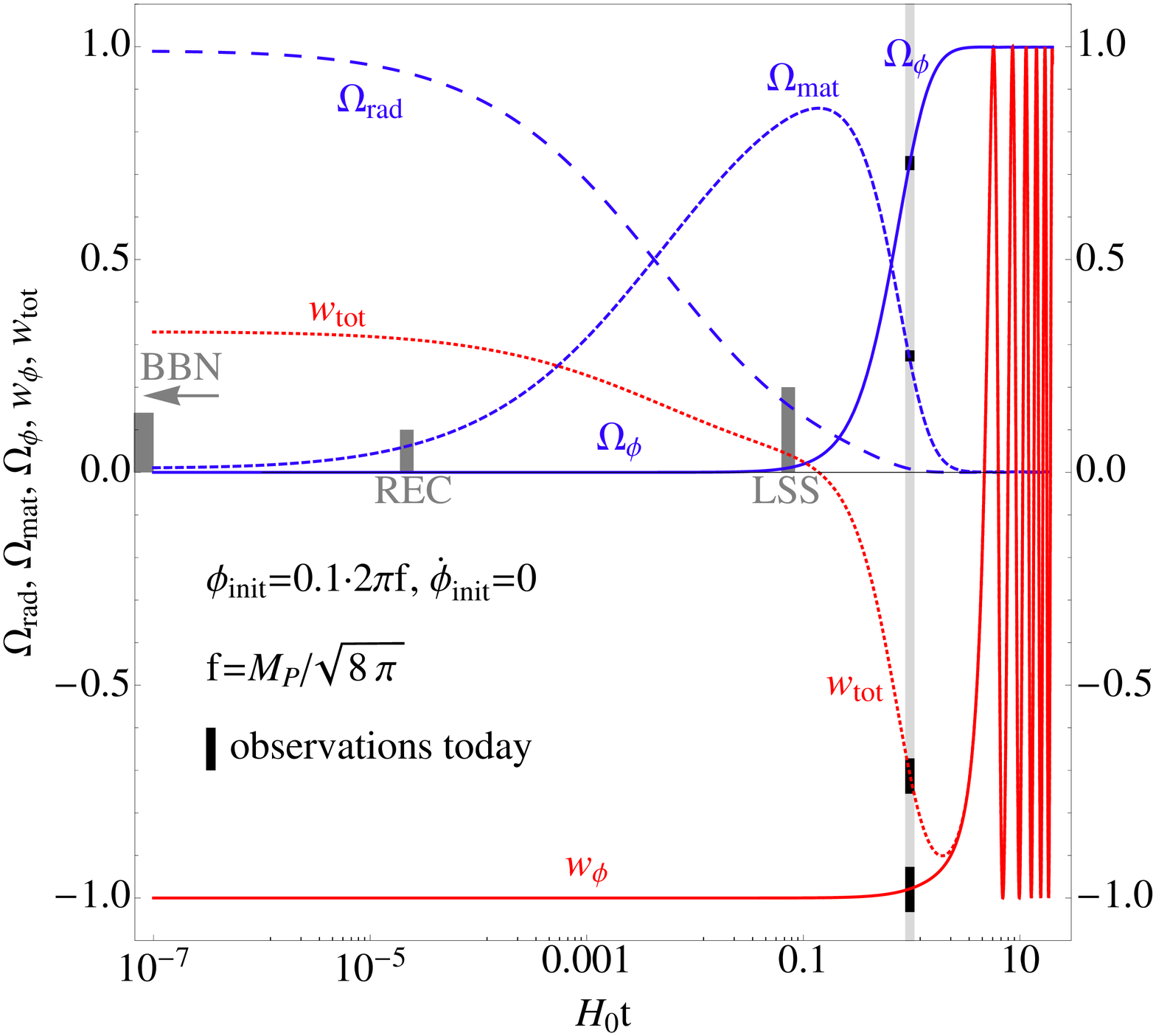}
\end{tabular}
\caption{\label{fig:twoplotsquintessence}Evolutions of the field $\phi$ (left panel) and of the energy density parameters for matter, radiation, and the quintessence field, as well as of the quintessence and total EoS parameters (right panel), including the constraints on $\Omega_\phi$ from big bang nucleosynthesis (BBN), recombination (REC), and structure formation (LSS).}
\end{figure}
The result of the analysis is displayed in Fig.~\ref{fig:twoplotsquintessence}: On the left, the evolution of the quintessence field $\phi$ is plotted as a function of time, whereas on the right the evolution of the whole Universe is displayed. The field trivially falls into its minimum and starts a damped oscillation around it. The evolution of all important energy densities and EoS parameters also behaves as expected. For early times, radiation remains dominant, which can be clearly seen from the total equation of state parameter $w_{\rm tot}$ that is close to $\frac{1}{3}$. Later on, matter starts to dominate and $w_{\rm tot}$ is pulled closer and closer to zero. The current time is marked by the grey line, which also indicates the current obervational bounds on the quantities under consideration. Dark Energy remains subdominant until shortly before today, but will later (when $w_{\rm tot}$ finally fuses with $w_{\phi}$) become the only component that matters. This also explains the oscillatory behavior of the EoS parameter: The field will have zero velocity, $\dot{\phi}=0$, at the turning points, which leads to $w_\phi=\frac{\frac{1}{2}\dot{\phi}^2+V_q}{\frac{1}{2}\dot{\phi}^2-V_q}=-1$, whereas at the minimum of the potential we would have $V_q=0$ and $w_\phi=+1$ accordingly.

\begin{figure}[t]
 \centering
 \begin{tabular}[h]{lr}
 \includegraphics[width=8cm]{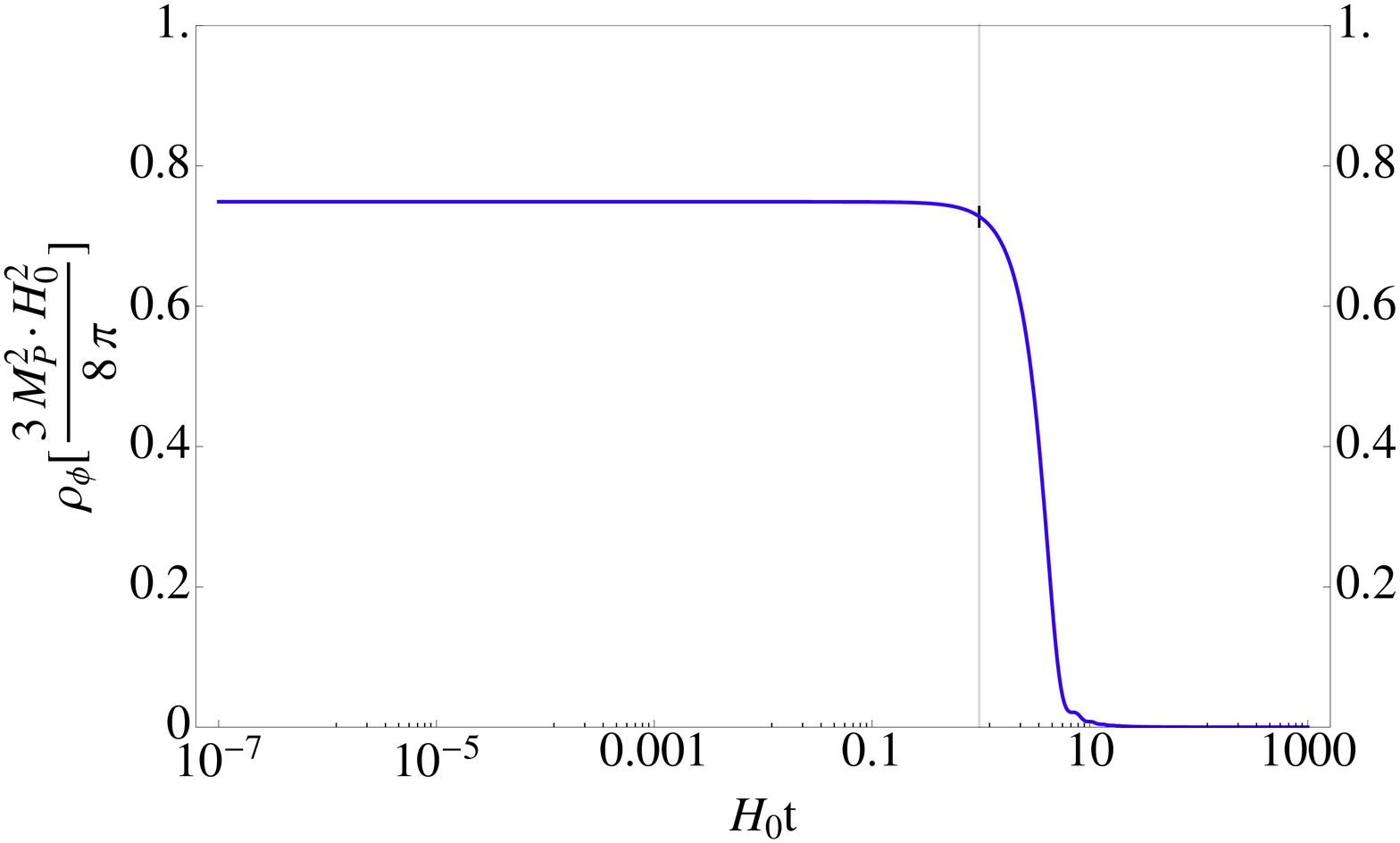} &
 \includegraphics[width=8cm]{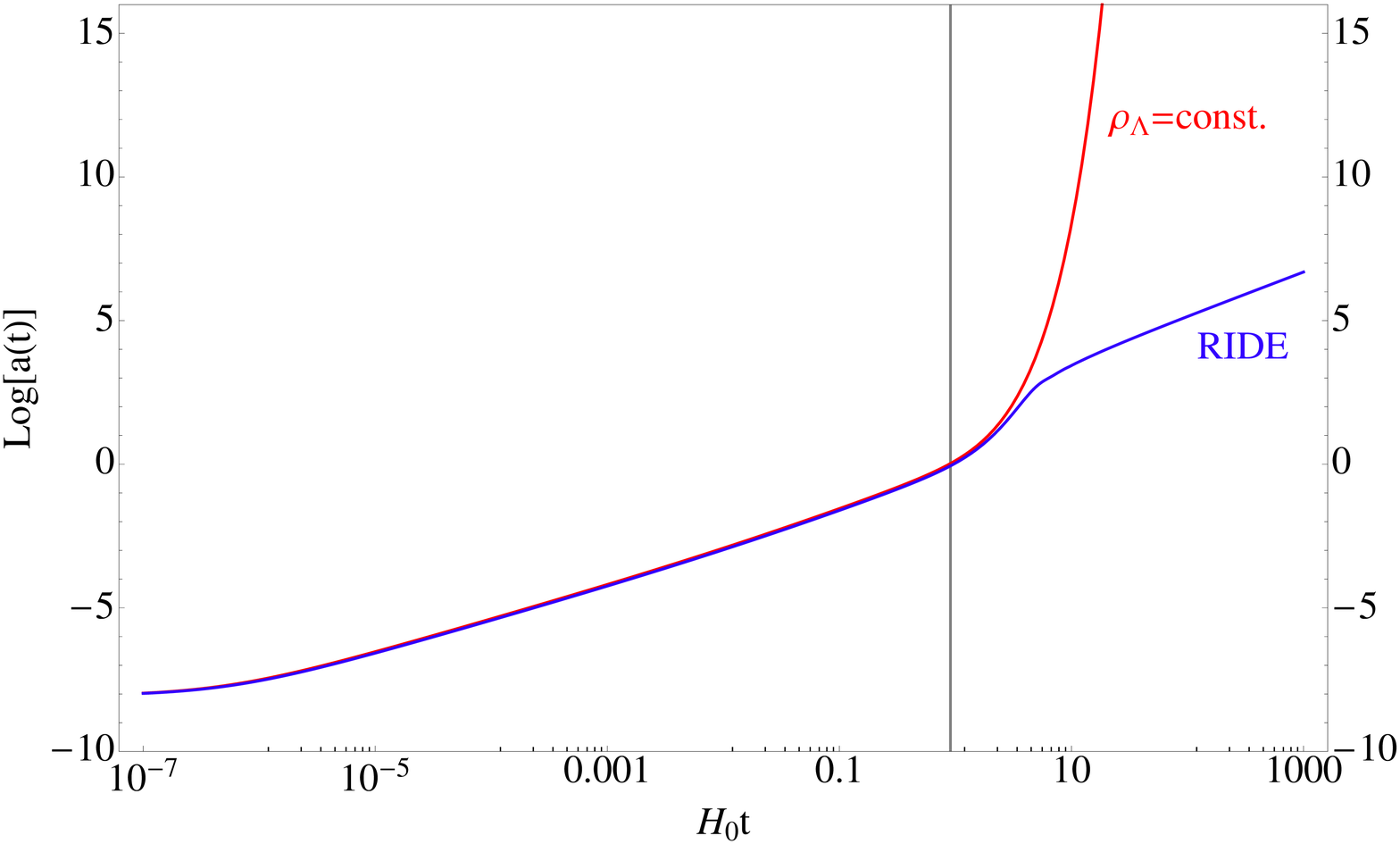}
 % DEdensity.eps: 139942656x0 pixel, 0dpi, infxnan cm, bb=
 \end{tabular}
 \caption{The evolution of the Dark Energy density $\rho_\phi(t)$ and the scale factor $a(t)$ in the RIDE model compared to a cosmological constant. We have explicitly verified that a different normalisation of the scale factor $a(t)$ does not change our results.}
 \label{fig:DEdensity}
\end{figure}

On the right panel, we have also indicated the bounds for early Dark Energy coming from big bang nucleosynthesis ($\Omega_{\Lambda}\lesssim 0.14$, \cite{Copeland:2006wr}), from recombination ($\Omega_{\Lambda}\lesssim 0.1$, \cite{garny}), and from structure formation ($\Omega_{\Lambda}\lesssim 0.2$, \cite{Doran:2001rw}), at times \cite{Kinney:2009vz} $t\sim3~\rm min$, $t\sim3\cdot10^5~{\rm y}$, and $t\sim 10^9~{\rm y}$, respectively, which all essentially indicate that Dark Energy should have become important only now. Note that the constraint arising from the formation of nuclei should actually be imposed at $H_0t\sim 10^{-15}$ \cite{Kinney:2009vz}, which is not displayed but indicated by the arrow in the right panel of Fig.~\ref{fig:twoplotsquintessence}. In the numerical analysis, we have normalised the evolution equations in such a way that the present Hubble constant $H_0$ equals one, and time is measured in inverse Hubble units. The age of the Universe is found to agree with $1/H_0$ within 5$\%$. The deviation from a cosmological constant becomes obvious in Fig.~\ref{fig:DEdensity}, where we have plotted the Dark Energy density on the left, which would simply be a horizontal line in the case of constant vacuum energy. The right panel shows the deviation of the scale factor from a scenario with a cosmological constant. The deviation at later times again comes from the behavior of the quintessence field $\phi$: It rolls down the potential towards its minimum and then performs a damped oscillation around that point (cf.\ right panel of Fig.~\ref{fig:potentials} and left panel of Fig.~\ref{fig:twoplotsquintessence}).

We would like to conclude this section, just noticing that the energy density at recombination
is predicted to be a standard mixture of matter and radiation.

%%%%%%%%%%%%%%%%%%%%%%%%%%%%%%%%%%%%%%%%%%%%%%%%%%%%%%%%%%%%%%%%%%%%%%
\section{\label{sec:SUGRAModel}An out of this world RIDE scenario}
%%%%%%%%%%%%%%%%%%%%%%%%%%%%%%%%%%%%%%%%%%%%%%%%%%%%%%%%%%%%%%%%%%%%%%

The remaining question is whether there are realistic scenarios that include our model. The example that we present here is a sequestered scenario which is consistent with the MSSM and has already been discussed in the literature in a similar fashion. It can be used as (toy) realization of RIDE.

Consider a superpotential,
\beq
W=W_{\rm obs} + W_{\rm seq},
\label{eq:W}
\eeq
where $W_{\rm obs}$ represents the observable sector, e.g. the MSSM spectrum, while $W_{\rm seq}$ represents a sequestered sector. Sequestered is more hidden than usual hidden sectors, since there will be no Planck scale suppressed operators that couple it to the observable sector. In practice this is achieved as described in Ref.~\cite{Randall:1998uk}, as we discuss now. The idea is that there are two 3-branes, an observable one and a sequestered one, separated by an extra dimension. We live on the observable brane, along with the MSSM particles, while our RIDE model lives on the sequestered brane.  Supersymmetry (SUSY) is badly broken in the sequestered sector, in our RIDE model, but the SUSY breaking is not easily transmitted to the observable sector, since the two branes are separated by the extra dimension coordinate, where the separation is sufficiently large and only gravity is in the bulk.
There are then no operators of order $1/M_P$ connecting the observable sector to the hidden sector, which is called ``sequestered''. So SUSY can be badly broken in the RIDE model without spoiling the observable MSSM.

However, as discussed in~\cite{Randall:1998uk} there will always be the gravity anomaly contribution to soft masses in the observable sector which gives soft masses $m_0 \sim \frac{m_{3/2}}{16\pi^2}$ due to a loop suppression, where $m_{3/2}$ is the gravitino mass. So we need to ensure $m_{3/2}\lesssim 100$ TeV.  Let us see how this could work in an example. The following example will also address the questions of the absence of the quartic scalar coupling and the origin of the radiative symmetry breaking of the scalar field.

We shall take $W_{\rm obs}=W_{\rm MSSM}$ for definiteness (although any SUSY model in the observable sector would suffice equally well) and the sequestered superpotential as follows:
\beq
W_{\rm seq}= M\Phi \overline{\Phi}+\lambda \Phi \psi \chi ,
\label{eq:Wseq}
\eeq
where $\Phi , \overline{\Phi}, \psi ,  \chi $ are independent superfield degrees of freedom, and we drop hats on superfields, which should not be confused with their scalar components.
The global SUSY $F$-terms include
\beq
F_{\overline{\Phi}}= M\Phi\ ,
\eeq
so that the potential includes a term
\beq
V=|F_{\overline{\Phi}}|^2= M^2\Phi^{\dagger}\Phi \ ,
\eeq
of the kind that we began with in Sec.~\ref{sec:Model}. {\rm Note that there is no quartic term in the potential, since in SUSY theories quartic terms arise from $D$-terms, and here the fields are supposed to carry no gauge charges.} Note that the second term in Eq.~\eqref{eq:Wseq} could, in principle, lead to dissipative effects~\cite{Berera:2009zz} in case it caused a decay of the quintessence field. However, we disregard this possibility here because of three reasons: First, the fields $\psi$ and $\chi$ carry no Standard Model charges and are barely coupled to any active fields (even the coupling to neutrinos could be easily switched off by a suitable symmetry). Second, these fields can be assumed to obtain very heavy masses such that they effectively decouple in the quintessence phase. And third because any treatment of a quintessence field decay is, in general, very model-dependent and beyond the scope of a toy model as the one presented in this paper. 

We now want the mass squared to be driven negative radiatively. This is achieved by the second term on the right-hand side of Eq.~\eqref{eq:Wseq} proportional to the Yukawa couplings $\lambda$. Loops of $\psi$ and $\chi$ will tend to drive $M^2$ negative in pretty much the same way as the Higgs mass squared is driven negative by top quark loops in the MSSM. The main difference is that here we need $M\sim 10^{11}$~GeV, and we require its square to be driven negative close to the Planck scale. In Appendix~\ref{app:RGE}, we show that large soft masses of $\psi, \chi$ (for example soft masses of order $10^{14}$ GeV in the considered example) are required to drive $M^2$ negative close to the Planck scale and result in a running scalar mass term of the kind that we parametrised in Eq.~\eqref{eq:inf_pot}. Such large soft masses are consistent with SUSY being broken in the hidden sector at very large scales as we now discuss.\footnote{Note that the presence of extra dimensions could significantly lower the estimated SUSY breaking scale as follows: Suppose that the sequestered brane has a number of extra dimensions which are parallel to it, as opposed to the extra dimension orthogonal to it which serves to separate it from the observable brane. The fields $\psi$ and $\chi$ feel these extra dimensions, while the field $\Phi$ does not, and their Kaluza-Klein (KK) excitations give rise to a large multiplicity of states which all enter the loop corrections of $M^2$, helping to drive it negative at a scale $\Lambda$ close to the Planck scale. The separation of the KK states depends on the size $R$ of the parallel extra dimension. Large $R$ corresponds to many different KK states within a given energy interval. Moreover, the number of KK states increases multiplicatively by adding more parallel extra dimensions. Such large multiplicity factors would serve to lower the above estimate of the SUSY breaking scale.}  

Assuming the radiative symmetry breaking mechanism just described, 
the VEV $\langle \Phi \rangle \sim \Lambda \sim 10^{-1}M_P$ results in a very large value for
\beq
\langle F_{\overline{\Phi}} \rangle \sim  M \Lambda \sim 10^{29} \ {\rm GeV}^{2} .
\label{eq:F}
\eeq
This large F-term VEV is also consistent with other F-term VEVs which are required to generate the large soft masses for 
the $\psi, \chi$ fields responsible for driving $\langle \Phi \rangle$ in the first place. 
Without the sequestering such large soft masses in the observable sector (far in excess of the TeV scale) would render the MSSM so badly broken as to be not relevant for the LHC, Dark Matter, the hierarchy problem, gauge unification, and so on. However, assuming sequestering, the observable sector soft masses may be at the TeV scale,
and the only requirement is that the gravitino mass does not exceed about 100~TeV, as discussed above. The gravitino mass arising from the sequestered sector is given by
\beq
m_{3/2} = e^{K/2M_P^2}\frac{\langle W_{\rm seq} \rangle }{M_P^2}  \sim \frac{\langle M \Phi \overline{\Phi} \rangle }{M_P^2}
\sim \frac{M\Lambda \langle \overline{\Phi} \rangle }{M_P^2} < 100 \ {\rm TeV}.
\label{eq:gravitino}
\eeq
Here $K$ denotes the K\"ahler potential which, in the canonical form, is just $\Phi^\dagger \Phi$. Since $\langle \Phi \rangle < M_P$, it is approximately correct to disregard the exponential. Inserting the value of Eq.~\eqref{eq:F} into Eq.~\eqref{eq:gravitino}, we find the constraint
\beq
\frac{ \langle \overline{\Phi} \rangle }{M_P} < 10^{-5}.
\label{eq:constraint}
\eeq
Since $\overline{\Phi}$ has no Yukawa couplings we would not expect it to have a radiatively driven VEV, so this constraint can easily be satisfied.

%%%%%%%%%%%%%%%%%%%%%%%%%%%%%%%%%%%%%%%%%%%%%%%%%%%%%%%%%%%%%%%%%%%%%%
\section{\label{sec:Conclusions}Conclusions}
%%%%%%%%%%%%%%%%%%%%%%%%%%%%%%%%%%%%%%%%%%%%%%%%%%%%%%%%%%%%%%%%%%%%%%

We have proposed a model based on radiative symmetry breaking that combines inflation with Dark Energy and is consistent with the WMAP 7-year regions. The RIDE model leads to the prediction of a spectral index $0.955 \lesssim n_S \lesssim 0.967$ and a tensor to scalar ratio $0.142 \lesssim r \lesssim 0.186$, both consistent with current data but testable by the Planck experiment. The radiative symmetry breaking close to the Planck scale gives rise to a pseudo Nambu-Goldstone boson with a gravitationally suppressed mass which can naturally play the role of a quintessence field responsible for Dark Energy. In the case of Dark Energy, the RIDE model predicts $w_{\phi}\neq-1$ at the present time ($w_\phi=-0.98$ in our numerical example), with the expansion of the Universe differing from the case of a cosmological constant in future epochs. Finally, we have presented an example scenario in which a RIDE toy model could arise. A next step of investigation could be to search for more realistic examples of RIDE, and to put them to a thorough test.

%%%%%%%%%%%%%%%%%%%%%%%%%%%%%%%%%%%%%%%%%%%%%%%%%%%%%%%%%%%%%%%%%%%%%%
\section*{\label{sec:Ack}\noindent{Acknowledgments}}
%%%%%%%%%%%%%%%%%%%%%%%%%%%%%%%%%%%%%%%%%%%%%%%%%%%%%%%%%%%%%%%%%%%%%%

We would like to thank M.~Garny and L.~Sorbo for providing useful information, as well as M.~Lindner for useful discussions. This work has been supported by the DFG-Sonderforschungsbereich Transregio 27 `Neutrinos and beyond -- Weakly interacting particles in Physics, Astrophysics and Cosmology'. The work of AM is supported by the Royal Institute of Technology (KTH), under the project no.\ SII-56510, , and by the G\"oran Gustafsson foundation. SFK and CL acknowledge support from the STFC Rolling Grant ST/G000557/1. SFK is grateful to the Royal Society for a Leverhulme Trust Senior Research Fellowship and a Travel Grant. PDB acknowledges financial support from the NExT Institute and SEPnet.

\appendix
%%%%%%%%%%%%%%%%%%%%%%%%%%%%%%%%%%%%%%%%%%%%%%%%%%%%%%%%%%%%%%%%%%%%%%
\section{\label{app:RGE}On the Renormalization Group Evolution}
%%%%%%%%%%%%%%%%%%%%%%%%%%%%%%%%%%%%%%%%%%%%%%%%%%%%%%%%%%%%%%%%%%%%%%
\label{app:facts}
\setcounter{equation}{0}
\renewcommand{\theequation}{A.\arabic{equation}}

From the superpotential in Eq.~\eqref{eq:Wseq}, it is easy to derive the SUSY-preserving Lagrangian in terms of component fields, supplemented by soft breaking terms~\cite{Martin:1997ns}. Using this, one can calculate the divergent part of the correction to the self-energy of $\tilde \eta$, just as for the MSSM Higgs case, which results in
\begin{equation}
 \Pi_{\tilde \eta \tilde \eta}^{\rm div} = \frac{4\lambda^2}{16 \pi^2} \left( m_b^2-m_f^2 \right) \frac{1}{\eps},
 \label{eq:self-energy}
\end{equation}
where $m_f$ ($m_b$) are essentially the masses of the fermionic (bosonic) components of the superfield $\Phi$, and where we have neglected the tri-linear coupling arising from soft breaking. Indeed this correction vanishes in the supersymmetric limit, $m_f=m_b$. Using the scaling of the 4-scalar coupling $\lambda$ in dimensional regularization, one can easily derive the corresponding renormalization group (RG) equation that describes the dependence of the mass square $m_{\tilde \eta}^2$ on the energy scale $\mu$~\cite{Covi:2004tp}:
\begin{equation}
 \frac{d(m_{\tilde \eta}^2)}{d \ln \tilde \eta}=\mu \frac{d(m_{\tilde \eta}^2)}{d\mu} = \frac{8\lambda^2}{16 \pi^2} \left( m_b^2-m_f^2 \right).
 \label{eq:RG-equation} 
\end{equation}
Indeed, the right-hand side of this equation has just the form expected for the general $\beta$-function of the scalar field under consideration (cf.\ Eq.~(3) in Ref.~\cite{Covi:2004tp}, where in our case $C=0$, due to the absence of gauge interactions, and $D=8$). Approximating the left-hand side by a difference quotient, one obtains
\begin{equation}
 m_{\tilde \eta}^2 (\mu=f) = m_{\rm tree}^2 + \frac{8 \lambda^2}{16 \pi^2} \Delta m_{\rm soft}^2 \ln \left( \frac{f}{M_P} \right),
 \label{eq:m2-function}
\end{equation}
where $m_{\rm tree}^2=M^2$, $\Delta m_{\rm soft}^2=m_b^2-m_f^2$, and $f=\sqrt{\frac{2}{e}} \Lambda$ is the VEV of $\tilde \eta$. Taking $\Lambda\sim 0.1 M_P$ (cf.\ Sec.~\ref{sec:Inflation}), $M_P=1.2\cdot 10^{19}$~GeV, and hence $M\sim 10^{11}~{\rm GeV}$ [cf.\ Eq.~\eqref{eq:F}], the requirement of the right-hand side of Eq.~\eqref{eq:m2-function} being negative results in the constraint
\begin{equation}
 \lambda \sqrt{\Delta m_{\rm soft}^2} \gsim 3\cdot 10^{11}~{\rm GeV},
 \label{eq:RG-constraint}
\end{equation}
and hence, e.g., $\sqrt{\Delta m_{\rm soft}^2}\sim 10^{14}~{\rm GeV}$ for $\lambda\sim 10^{-3}$, which is indeed far above the TeV scale, as anticipated.

%\newpage
%=============================================================================
\bibliographystyle{./apsrev}
\bibliography{./RIDE}
%=============================================================================

\end{document}